\documentclass[preprint]{aastex}

\begin{document}

\title{The M31 Globular Cluster Luminosity Function\footnote{
Based partially on observations made with the NASA/ESA Hubble Space Telescope,
obtained from the data archive at Space Telescope Science Institute. 
STScI is operated by the Association of Universities
for Research in Astronomy, Inc. under NASA contract NAS 5-26555.}
}

\author{Pauline Barmby\footnote{Guest User, Canadian Astronomy Data Centre, 
which is operated by the Herzberg Institute of Astrophysics, National Research Council of Canada.}
 \& John P. Huchra}
\affil{Harvard-Smithsonian Center for Astrophysics, 60 Garden St., Cambridge, MA 02138} 
\email{pbarmby@cfa.harvard.edu, huchra@cfa.harvard.edu}
\and
\author{Jean P. Brodie}
\affil{Lick Observatory, University of California, Santa Cruz, CA 95064}
\email{brodie@ucolick.org}

\shortauthors{Barmby, Huchra, \& Brodie}
\shorttitle{M31 GCLF}

\begin{abstract}

We combine our compilation of photometry of M31 globular clusters and 
probable cluster candidates with new near-infrared photometry for 30 objects. 
Using these data we determine the globular cluster
luminosity function (GCLF) in multiple filters for the M31 halo clusters.
We find a GCLF peak and dispersion $V_0^0=16.84\pm0.11$, ${\sigma}_t=0.93\pm0.13$
(Gaussian ${\sigma}=1.20\pm0.14$), consistent with previous results. 
The halo GCLF peak colors (e.g. $B_0^0-V_0^0$) are consistent with 
the average cluster colors. We also measure $V$-band GCLF parameters for 
several other subsamples of the M31 globular cluster population.
The inner third of the clusters have a GCLF peak 
significantly brighter than that of the outer clusters
($\Delta V^0\approx 0.5$~mag).
Dividing the sample by both galactocentric distance and metallicity,
we find that the GCLF also varies with metallicity, as 
the metal-poor clusters are on average 0.36~mag fainter 
than the metal-rich clusters. Our modeling of the catalog 
selection effects suggests that they are not the cause of the
measured GCLF differences, but a more-complete, less-contaminated
M31 cluster catalog is required
for confirmation. Our results imply that dynamical destruction 
is not the only factor causing variation in the M31 GCLF:
metallicity, age, and cluster initial mass function  may also
be important.

\end{abstract}

\keywords{galaxies: individual (M31) -- galaxies: star clusters -- 
  globular clusters: general}

\section{Introduction}

Globular clusters (GCs) are unique markers of the formation
and evolution of galaxies. They are bright, easily identifiable
packages of Population~II stars with homogeneous abundances 
and history. GCs are relics of the earliest star formation
in galaxies and have witnessed many changes in their
parent galaxies. The study of globular cluster systems (GCSs) 
is thus important in understanding the history of galaxies.

The distribution of integrated GC magnitudes, known as the
globular cluster luminosity function (GCLF), 
has long been known to be unimodal and approximately symmetric
in the Milky Way. 
The assumption that these properties are universal 
has allowed the determination of GCLF parameters for
over a hundred other galaxies, and the peak absolute magnitude 
is found to be roughly constant from galaxy to galaxy \citep{har88}. 
Since the variation in mass-to-light ratios among GCs
is fairly small \citep[see, e.g.,][]{dg97} the constant 
peak magnitude implies the existence of a characteristic mass
scale. Theorists have attempted to explain why there should be a 
characteristic mass scale for globular clusters: is it a 
property of formation \citep[e.g.][]{pd68}, a result of subsequent dynamical
processes \citep[e.g.][]{ot95}, or a combination? The constant peak magnitude
also presents a challenge to observers, who have attempted to 
quantify the variation in peak magnitude by environment \citep{bt96},
galaxy type and luminosity \citep{whi97}, and color \citep{kws99}. 
Many observers have also attempted to use the GCLF peak
as a standard candle for distance measurement. This method
has had a mixed reception, with some authors \citep[e.g.][]{whi97} 
claiming good results and others \citep[e.g.][]{fer99} less complimentary.

GCLF measurements are made in a number of different bandpasses.
The most widely used are $V$ and $B$, (and their Hubble Space Telescope 
(HST) equivalents, F555W and F450W); the I-band equivalent F814W is
also commonly used for HST data.
$B$ and $V$ band mass-to-light ratios are sensitive to metallicity,
so GCSs with the same underlying mass distribution but 
different average metallicities will have different GCLFs.
\citet*{acz95} used the relation between metallicity and
mass-to-light ratio as predicted by \citeauthor{w94}'s (1994)
simple stellar population models to estimate the effects of
metallicity variations on the GCLF peak. Metal-rich clusters should
be fainter (their $M/L$ is larger), and the effect on the GCLF can
be substantial. For example, \citet{acz95} predict that a change in mean GCS
metallicity from $\overline{\rm [Fe/H]}=-1.35$~dex (the Milky Way value) to 
$-0.60$~dex \citep[the value for the elliptical NGC~3923;][]{z95}
shifts the GCLF peak by $\Delta B^0=0.35$ and $\Delta V^0=0.22$.
The same metallicity change shifts the $J$ band GCLF peak by $\Delta J^0=-0.09$,
because the metallicity effect on $M/L$ changes direction for bandpasses
redward of $I$. Observing at longer wavelengths also reduces the effects 
of extinction, both Galactic and within the GCS parent galaxy.

Destruction of globular clusters through dynamical effects
could produce a globular cluster mass function (GCMF) that varies
with distance from the center of the host galaxy. 
The three effects usually considered by modelers
are disk shocking, evaporation, and dynamical friction;
the first two destroy low-mass, low-concentration clusters, while
the last is most effective for massive clusters. All three effects are 
strongest at small Galactocentric distance $R_{gc}$. Even at small $R_{gc}$, the 
dynamical friction timescale for typical-mass GCs is much longer than
a Hubble time, and dynamical friction is probably not important
in GCMF evolution \citep{ot95}. 
Low-mass clusters near the Galactic center are thus 
most prone to destruction.
Assuming a mass-to-light ratio independent of $R_{gc}$
-- if the GCS has a radial metallicity gradient this is not
strictly true since $M/L$ depends on [Fe/H]
-- the radial difference in the GCMF can be
translated into a radial difference in the GCLF. 
Many authors \citep{b97,og97,v98} have predicted the size
of the GCLF variation, with widely differing results 
(see Section~\ref{sec-implic}).

\begin{figure}
\includegraphics[scale=0.7]{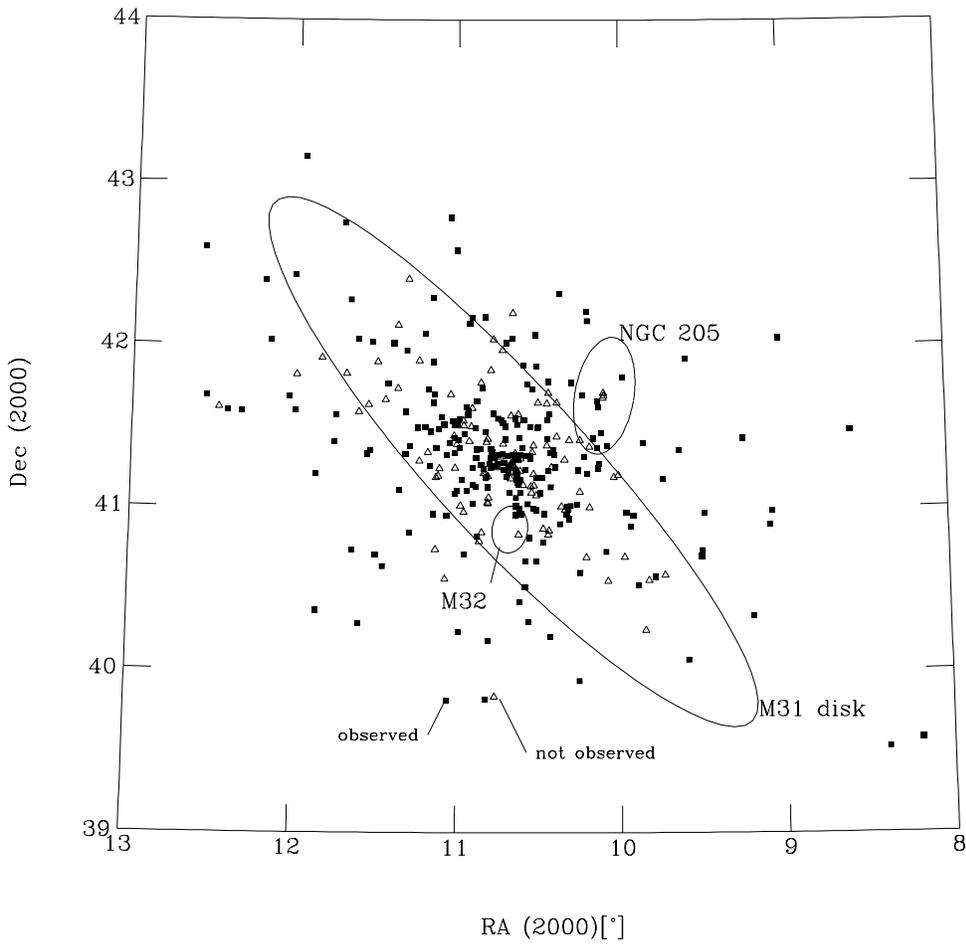}
\caption
{Photometric completeness for IR photometry (at least one of $J, H$, or $K$) of M31 GCs.
Symbols indicate whether an object has been observed.\label{magcomp-ir}}

\end{figure}

In \citet{b00} we compiled the best available photometry
for 435 M31 globular clusters and plausible candidates.
This catalog contains $V$ magnitudes for almost all
objects and $B$ magnitudes for about 90\%, but completeness
in the longer-wavelength bands is much lower. About 
55\% of the objects have $I$ magnitudes, and the same fraction,
although not the same objects, have $JK$; Figure~\ref{magcomp-ir}
shows the completeness of the existing IR photometry.
In this paper we estimate the parameters of the halo
clusters' GCLF in six different bandpasses; this includes
the first GCLF measurements in the near-IR.
We then investigate the variation in the $V$-band GCLF parameters for several
different subsamples of M31 GCs. Finally, we consider the implications
of the measured GCLF variations for GCS and galaxy formation and evolution.

\section{New observations and data reduction}
We obtained new near-IR photometry for M31 GCs and candidates
using two telescope/instrument combinations: Gemini \citep{gem} 
on the Lick Observatory 3-m telescope, and Stelircam \citep{tol98} 
on the F.~L. Whipple Observatory 1.2-m.
The Lick observations (in $J$, $H$ and $K^{\prime}$) 
were made on 1999 October 15-17 and the
Whipple observations (in $J$ and $K$) on 1999 December 26. 
Total integration times 
ranged from 90 seconds (re-observations of bright clusters in $J$ 
with Gemini) to 1900 seconds (observations of the faintest halo
clusters with Stelircam), using 4 to 9 dithered frames per object.

We used almost the same data reduction procedure for 
both cameras, which consisted of bias-subtraction (Gemini only),
dark-frame subtraction, non-linearity correction (Stelircam only),
flat-fielding (using a `superflat' made from 50-100 blank-sky frames), 
sky-subtraction, registration, and co-addition.
Two methods of sky-subtraction were used: for objects on
the galaxy disk we made separate sky observations and subtracted
their median, and for objects without galaxy background we subtracted
a `running sky' made from the median of 4-8 temporally contiguous 
frames. 
We measured large-aperture magnitudes for \citet{eli82} and \citet{per98} 
standard stars, typically 10-15 stars per filter per night, and
fit a two-component (zeropoint and airmass coefficient)
photometric solution using standard star magnitudes.
Residuals of the photometric solution were typically about $0.01-0.02$~mag.
The difference between $K$ and $K^{\prime}$ magnitudes
of standard stars is smaller than these residuals (M.~Pahre, 2000, private
communication), so we assume that our $K^{\prime}$ measurements adequately
represent the $K$-band.

Our previous aperture photometry of globular clusters used 12-arcsec 
diameter apertures, and we found that, for the Gemini and Stelircam 
data, this aperture again  contained about 95\% of the total cluster light. 
The new measurements appear in Table~\ref{newphot}. Some of these data
are for objects whose previous IR photometry disagreed strongly
with their spectroscopic metallicities \citep[mostly blue clusters in][]{b00}; 
our new measurements yielded colors more consistent with the
spectroscopic data. The original goal 
of these observations was to complete the IR photometry 
for the M31 halo sample of \citet*{rhh94}.
Poor observing conditions prevented us from completing 
the halo sample but we did obtain new photometry for 38 objects in total.
The total number of M31 clusters
and cluster candidates with $JK$ photometry is now $\sim 250$, including
the 30 objects measured for the first time in this paper.

\section{Preliminaries to the GCLF}

\subsection{Sample definitions\label{sampdef}}

The standard approach to correcting for incompleteness
and contamination in the GCLF \citep[see, e.g.,][]{for96b} 
is not feasible for M31. Our
catalog is a compilation of several previous cluster searches,
each with different sky coverage and selection functions.
Foreground contamination is a more severe problem for
M31 than for more distant galaxies: because M31 is nearby, 
it covers a large angular area and its GCs have
apparent magnitudes comparable to those of Galactic stars. 
A list of `failed' M31 GC candidates in \citet{b00} 
shows that about 15\% are foreground stars.
Other possible contaminants are background galaxies 
and objects such as open clusters and H~II regions
in the galaxy itself.
Spectroscopy or high-resolution imaging are the only ways to 
unambiguously discriminate M31 GCs from contaminating
objects, but these techniques have been used mostly on the
brighter clusters (see Section~\ref{contam}).
\citet{bat87}, the major source of clusters and candidates for our
catalog, estimate that 80-90\% of their probable (class A and B)
cluster candidates with $V<18$ are true M31 GCs. 

Defining a complete, uncontaminated sample of M31 clusters
to a magnitude limit well past the GCLF turnover is not possible
with currently available data. There is no quantitative estimate
of either the completeness or contamination rate, and any attempt
to estimate these from the confirmed clusters is 
hampered by the strong dependence of both completeness and contamination
on apparent magnitude and position in the galaxy. 
To reduce these effects as much as possible
while still retaining a reasonable sample size, we define our `full
sample' to be all clusters in the \citet{b00} catalog with $V<18$,
$B-V>0.45$, and $R>5$\arcmin, where $R_{gc}$ is the projected distance
from the center of M31.
The magnitude criterion restricts the sample to the  
range where incompleteness and contamination should be minimal.
The color criterion is needed to remove objects in M31 which are
bluer than any Galactic globular cluster and may possibly be 
younger clusters. We use here a slightly less stringent color 
criterion than in \citet{b00} in order to retain as many of the 
halo clusters as possible; many of them are slightly ($\sim0.02$~mag)
bluer than the previous limit of $B-V=0.55$.
Restricting the sample to $R>5$\arcmin\ also
aims to reduce the incompleteness and contamination, as
the bright galaxy background near the M31 bulge
makes cluster identification and confirmation more
difficult. Several groups (\citealt*{wsb85,ach92}; \citealt{bat93}) have attempted to
define a complete sample of GCs near the M31 nucleus.
However, the published catalogs are likely to be both
contaminated and incomplete: \citet{b00} showed that many 
of the objects listed by \citet{bat93} are not globular clusters, 
and \citet{mos99} recently identified 33 previously 
unknown GC candidates in the M31 bulge.

Our full sample has a total of 294 clusters and
cluster candidates. The overlap with \citeauthor{bat93}'s
(\citeyear{bat93}) `adopted best sample' is about 85\%;
the differences are largely due to our exclusion of
GCs with $R_{gc}<5$\arcmin\ and \citeauthor{bat93}'s use of
a red color limit.
To investigate the effects of metallicity on the GCLF, we
define metal-poor (MP) and metal-rich (MR) subsamples by splitting the
full sample at ${\rm [Fe/H]}=-1.0$. Metallicities for most objects
without spectroscopic data were estimated from colors as in \cite{b00};
we accounted for clusters without metallicities by modifying
the completeness functions.
The metal-rich subsample has 75 objects and the metal-poor has 191.
For comparison with the halo sample (see below), 
we define a disk sample consisting of all objects in the full 
sample which are not members of the halo.
There are a total of 226 `disk' objects (the number of 
disk plus halo objects is greater than the number of objects
in the full sample because 18 halo objects have $V>18$).

Measuring the GCLF of halo GCs in M31 avoids the problem of
having to determine individual extinctions for each cluster,
since these clusters should be subject to only foreground extinction.
\citet{rac91} began a concerted effort to define 
a sample of GCs in the M31 halo that was complete to a faint magnitude
limit ($V\approx19$) and free of contamination. He defined the halo boundary as
a 260\arcmin$\times56$\arcmin\ ellipse with position angle $38\arcdeg$, centered on the
M31 nucleus (see Figure~\ref{magcomp-ir}).
The effort to produce a complete, uncontaminated halo sample was continued 
by \citet*{rhh92}, \citet{rac92} and \citet{rhh94}, and 
culminated in the published GCLFs of \citet{sec92}, 
\citet[hereafter SH93]{sh93}, and \citet{rhh94}.
Because of this work, incompleteness and contamination are much less severe 
for the halo clusters than for the overall population of GCs in M31.
To construct our halo sample, we start with
the sample of \citet{rhh94}. We remove the following three objects
(see \citet{b00} for our naming convention) as probable non-clusters:
\begin{itemize}
\item 168D-D020: ``probably not a cluster'' \citep{bat87} 
\item 109D-000: suspected to be foreground star by both \citet{bat87} and \citet{rhh94}
\item 007-059: shown spectroscopically to be a foreground star \citep{b00} 
\end{itemize}

Some of the GCs around M31 are near its companion, the dwarf elliptical
NGC~205; the clusters' formation and evolution may have been influenced 
by NGC~205's presence as well as that of M31.
There is some evidence that the GCLF peak absolute magnitude in 
dE  galaxies is fainter than that in large galaxies 
\citep{dur96,h00}, so we attempt to remove these `contaminating'
objects from the M31 halo sample.
We follow \cite{rhh92} in considering 011-063, 324-051, 331-057, and 330-056
to be `NGC~205 clusters' and dropping them from the M31 halo.
We also remove three additional clusters from the M31 halo list:
\begin{itemize}
\item 328-054: very close to NGC~205 in both position (1.2\arcmin)
and radial velocity (18~km~s$^{-1}$)
\item 326-000, 333-000: both $<2$\arcmin\ (400~pc) from NGC~205. No 
radial velocity information is available for these objects, and
their projection onto NGC~205 also makes accurate photometry difficult.
\end{itemize}

We add 227-000 to our list, although \citet{rhh92} reject this object from 
their halo sample because of its previous photometry. The rejection appears to 
have been 
a mistake, since their observations of this object put it within their color 
and curve-of-growth acceptance criteria. Our final halo list has 86 objects.
The list used by \citet{sec92} has 82 objects, as he excludes some `low-probability'
objects (469-220, DAO054, DAO055, DAO094, 456-000) while we retain these
for the sake of completeness.
The faintest object in our halo sample has $V=18.98$, and we believe
the sample to be essentially complete.
Our catalog is probably missing a few distant, Palomar-type
clusters in the M31 outer halo, but no M31 GC survey yet performed
would have revealed such objects. (Finding  Palomar-type GCs in M31 
would require surveys to $V\approx22$, in excellent seeing, to
a radius of 5 degrees from M31.) In any case, the luminosity
function should be little affected by the presence or absence of
a few clusters far from the LF peak.

\subsection{Extinction correction\label{extcorr}}

One of the major problems with using the Milky Way and M31 globular
cluster systems to calibrate the GCLF is the uncertain effect of extinction.
Many of the Galactic GCs are in the Galactic plane and are
highly extinguished; however, at least for those clusters the 
extinction can be determined from the color-magnitude diagram (CMD)
and spectroscopy of individual stars. In M31, globular clusters'
CMDs are not precise enough for determination of extinction, and only 
about a dozen clusters have published CMDs. Instead, measurements of
extinction for individual clusters \citep{vet62b,cra85,sl89,b00} have
relied on correlations of spectroscopic metallicity and/or
reddening-free parameters with intrinsic color. This procedure 
necessarily requires a calibration from the Galactic clusters
and the assumption of similar extinction laws, but \citet{b00}
showed that both of the these steps were reasonable.

The first step in determining the extinction of the M31 GCs
is to decide on a value for the foreground extinction.
The H~I maps of \citet{bh84} and the DIRBE/IRAS maps
of \citet*{sfd98} agree that the Galactic reddening 
in the direction of M31 is $E(B-V)=0.08$, corresponding to
$A_V=0.25$. In \citet{b00}, we found that
the average reddening of the M31 halo GCs is consistent with 
this value, although the individual reddening values had 
a large scatter. Since about half the
halo clusters do not have reliable reddening values, 
and about 20\% have no reddening determination at all,
we choose to use a single value, the foreground reddening,
for all the halo clusters. A few halo objects had well-determined
reddening values that were higher than this, but adopting
a single value for all objects simplifies comparison
of our results with those of other authors.

Adopting a single extinction value for the M31 disk clusters
is, of course, not reasonable. Fortunately, most of the disk objects 
in our GCLF sample (161/226) have reliable reddening values
determined in \citet{b00}. 
A few of the reliable values are actually below the foreground
value, and we replaced them with the foreground value. 
Some of the objects without reliable reddening had a 
high value of ${\sigma}_{E(B-V)}$. We examined the values
of $E(B-V)$ from individual colors for these objects, and found that
removing one or two discrepant values usually yielded an
acceptable error in the resulting reddening value. This left
42 objects without reddening values. For these objects we
estimated the reddening by computing the average reddening of
objects projected within 10\arcmin, weighted by the inverse of
the projected distance. This procedure is somewhat arbitrary,
and may yield erroneous results if the `nearby' clusters
are at significantly different distances along the line of sight.
A test of the technique on GCs with measured reddening values
produced reassuring results: the average offset between 
measured and estimated reddening values was $0.00\pm0.01$~mag,
although the scatter in the offset was large (0.19~mag).
We compute extinction values from $E(B-V)$
using the curve for $R_V=3.1$ given in \citet*{ccm89}.

\subsection{Completeness correction\label{compcorr}}

While we do not have a quantitative estimate of the magnitude
of completeness and contamination in our catalog as a whole,
there are two effects for which we can correct. One effect
is incompleteness caused by our magnitude limit.
This is somewhat complicated because we use an observational
(i.e. uncorrected for extinction) magnitude limit
but we need to determine the completeness function in 
extinction-corrected magnitudes, where we measure the GCLF. 
For example,
we believe our sample of all M31 clusters is complete to $V_{\lim}=18$, but
this does not mean it is complete to $V_0=17.75=18-A_V(\min)$. 
Heavily-extinguished clusters would have $V>18$ but $V_0<17.75$.
We compute the required completeness correction in $V_0$ bins as follows.
First, we divide each bin into five smaller sub-bins. 
For each sub-bin, we use the observed distribution of extinction 
values to determine the fraction of the bin which could have been
observed, i.e. for which $V_0 + A_V<V_{\rm lim}$. This fraction is the
completeness fraction, which we average over the sub-bins to determine
the completeness for each bin. We use the same procedure to correct
for the effects of the magnitude limit in $V$ on 
data in other bandpasses (written generically as $X_0$). 
In this case both the color and extinction
distributions are taken into account, and the objects which could have 
been observed have $X_0 + (V-X_0) = X_0+(V-X)_0+ A_V < V_{\rm lim}$.  
We compute individual completeness functions for all of the different
M31 GC samples discussed in this paper.

We performed Monte Carlo simulations to check that this correction procedure
did not introduce a bias in the GCLF results. We generated $V_0$ GCLFs from 
Gaussian and $t_5$ \citep{sec92} distributions with $V_0=17.0$, 
${\sigma}_g=1.1$, ${\sigma}_t=0.9$, $N_{gc}=300$,
parameters appropriate to the M31 GCLF. We applied to each
object an extinction chosen randomly from the observed distribution, 
removed the objects with $V>V_{\lim}$, generated the completeness function 
using the remaining objects, and measured the GCLF parameters of the sample. 
We corrected the GCLF parameters for the magnitude limit of each of the 
200 trials using the method described in the following section.
Over all trials, the averages of the peak location measure were $16.97\pm0.01$ for
Gaussian and $17.04\pm0.01$ for the $t_5$ distribution; the average
measures of dispersion were ${\sigma}_g=1.10\pm0.01$, ${\sigma}_t=0.92\pm0.01$.    
The uncertainties in the 200 individual measurements of peak location and
dispersion ($\sim0.18$ and 0.13~mag, respectively) are large compared to 
the average difference between
input (simulated) and output (measured) GCLF parameters,
so we conclude that our completeness correction does not introduce
a significant bias.

Another kind of incompleteness we correct for is missing photometry.
We have $V$ magnitudes for all of our objects, but observations
in other filters are not always available. Estimating the photometric 
completeness in bands other than $V$ is a simple procedure:
\begin{enumerate}
\item Set up a series of bins in magnitude $X_0$.
\item Sort objects with known $X_0$ magnitude into the appropriate bin.
\item Estimate the $X_0$ magnitudes of the other objects as $V-(V-X_0)$,
where $(V-X_0)$ is drawn at random from the probability distribution of the
objects with $X_0$, and sort them into the appropriate bins.
\item Add the bin totals from the previous two steps.
\item For each bin, the photometric completeness is (number of objects
with measured $X_0$)$/$(total number of objects in bin).
\end{enumerate}
The final completeness function is the average of 500 trials of
steps 3--5, in order to average out the
noise produced by drawing colors at random from the observed color distribution.
The completeness function from missing photometry is multiplied
by the completeness function for the magnitude limit to produce the
final estimate of completeness in each bandpass. 

We also checked the effects of errors in photometry and reddening 
on the GCLF results. Photometric uncertainties for the CCD and photoelectric 
photometry in our catalog are generally $\lesssim0.05$~mag;
uncertainties for photographic photometry (which is relatively uncommon
in our catalog) are larger, perhaps 0.15~mag. These uncertainties
are smaller than the errors in individual extinctions: in \citet{b00}
we estimated uncertainties in $E(B-V)$ to be $0.05-0.10$~mag,  
corresponding to $A_V$ uncertainties of $0.16-0.31$~mag. 
Using the same Monte Carlo simulations described above, we added 
Gaussian errors (from distributions with zero mean 
and standard deviations from 0.15--0.50~mag)
to every individual extinction in each of the 200 GCLFs. 
If an `extinction-with-error' value fell below the foreground value of 0.25,
the extinction was changed to the foreground value as in our
observational procedure. We also added errors to the measured
magnitudes, again drawn from Gaussian distributions with 
zero mean and dispersion increasing as $10^{0.2m}$ (the behavior
expected from photon statistics). 

Even large (${\sigma}_{A_V}=0.50$~mag) extinction errors did not 
change the average peak and dispersion measured for the GCLF ensemble 
by more than about 0.03~mag. Little change in the peak is expected,
since the extinction errors have zero mean and even an error of
0.5~mag does not shift many observed magnitudes beyond the magnitude limit.
The small effect of ${\sigma}_{A_V}$ on the measured GCLF dispersion 
(${\sigma}_g\approx 1.1$~mag) is explained by noting that
the two dispersions combine in quadrature 
and, for large values of ${\sigma}_{A_V}$, 
the magnitude limit truncates the observed distribution width.
Large magnitude errors (${\sigma}_m \gtrsim 0.3$ at $V=17$,
much larger than our estimated uncertainties)
biased the average GCLF peaks to brighter values but did not
significantly affect the dispersions. These effects are not
unexpected: large magnitude errors push more objects
beyond the magnitude limit (resulting in a brighter peak) 
and the magnitude again limit truncates the observed distribution width.
We conclude that extinction errors and magnitude uncertainties do not 
significantly affect our measurement of the GCLF parameters.

\section{The GCLF}

To compute the globular cluster luminosity functions, we used
version 2.01 of the maximum-likelihood code described by SH93.
This program estimates the maximum-likelihood values of 
the parameters $\hat{\Theta}=({\hat{m}}^0,\hat{\sigma})$ 
by finding the maximum in the likelihood function
\begin{equation}
L(\Theta)=\log l(\Theta) = \sum_{i=1}^N \log [\phi(m_i)].
\end{equation}
The simultaneous distribution function $\phi(m)$ is 
evaluated for each of the $N$ magnitude observations $m_i$:
\begin{equation}
\phi(m)=I(m)\gamma(m)
\end{equation}
$I(m)$ is the completeness function and $\gamma(m)$ is the intrinsic
GCLF --- a Gaussian or $t_5$ distribution with center $m^0$ and 
dispersion $\sigma$.
This is a simplified version of SH93's $\phi(m)$; 
we set the number of background objects to zero as we do not have
a quantitative estimate of the contamination rate. 
We also do not use the option to convolve the GCLF with the photometric error 
distribution: our photometry and extinction errors
are not well-defined functions of magnitude, and trials 
using rough estimates of the error distributions showed that
this did not significantly change the results.
Uncertainties in the parameters are determined by computing $l(\Theta)$
over a grid in parameter space centered on the most probable parameter
values. The resulting grid of likelihood values is normalized
to have a total probability of unity and collapsed along the parameter axes;
the 1$\sigma$ ranges for the parameter estimates are the ranges 
containing 0.68 of the total probability along each axis.

SH93 show that the results of their maximum likelihood procedure
are biased by the existence of a magnitude cutoff. The brighter the
limiting magnitude, the more the parameter estimates are biased toward
brighter $m^0$ and lower $\sigma$. To correct for
this effect, we use the results of their simulations, shown in 
their Figure~3.\footnote{We note what appears to be a misprint in SH93:
in both Figure~3 and the text, $(m^0-m_l)$ appears where $(m_l-m^0)$ is clearly implied.}
The equation used is:
\begin{equation}
\hat{m}^0=m^0+\delta(m_l-m^0)
\end{equation}
where $\hat{m}^0$ is the estimated value of the turnover produced
by the ML method, $m^0$ is the true, unbiased value, $m_l$
is the limiting magnitude, and $\delta$ is the function plotted in
the top panel of Figure~3 in SH93. We fit a fourth-order
polynomial to $\delta$, and use it to solve for $m^0$ given
$\hat{m}^0$ and $m_l$. We fit a different polynomial $\tilde{\delta}$
to $\hat{\sigma}=\sigma+\tilde{\delta}(m_l-m^0)$ and compute
$\sigma$ directly after
computing $m^0$.

\subsection{Results\label{sec-res}}
\begin{figure}
\includegraphics[scale=0.7]{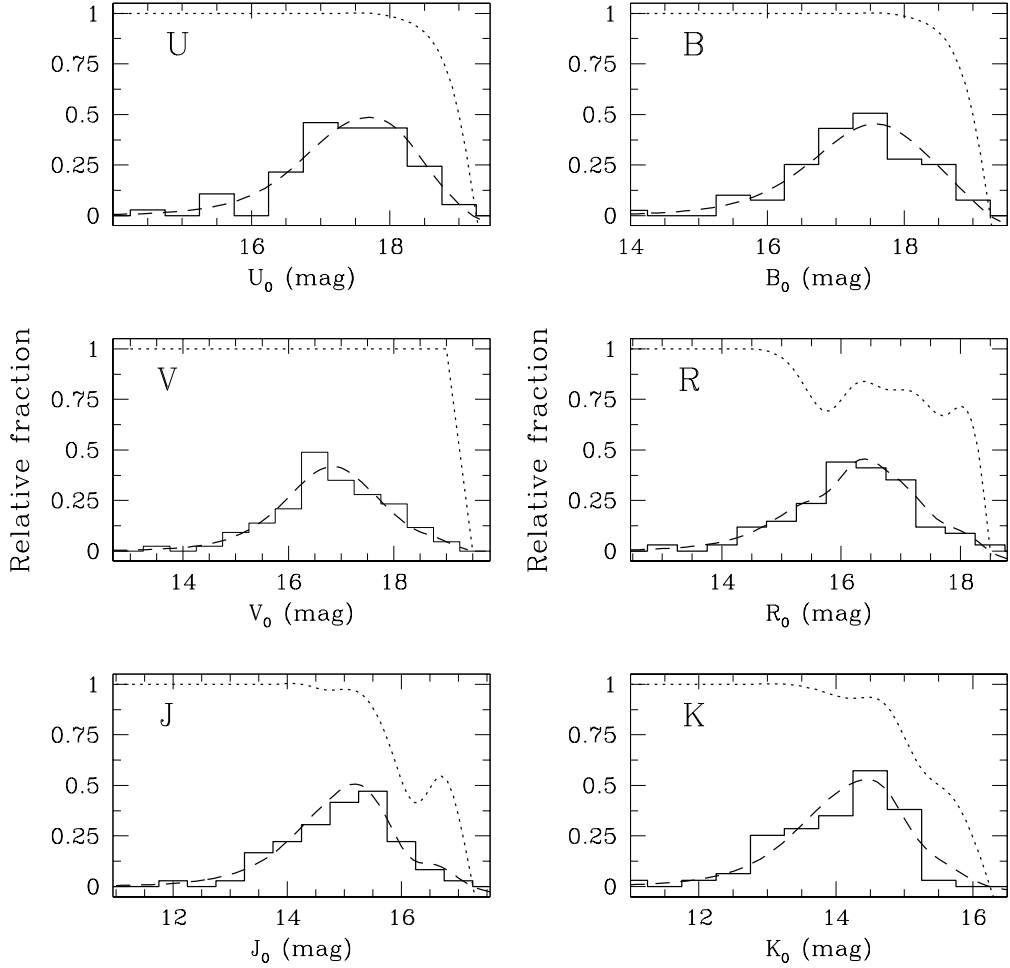}
\caption{GCLFs and completeness functions for M31 halo clusters. The dip in the
$R$-band completeness function is real; five clusters with $16.25<V<16.75$
(estimated $15.5<R_0<16.0$) do not have measured $R$ magnitudes.
\label{halo-fig}}
\end{figure}

We discuss the halo sample first, because most of the recent determinations
of the M31 GCLF have been for this sample. Since the halo sample is close
to complete, we can also estimate the GCLFs in other bandpasses
without correcting for a $V$ magnitude limit.
Figure~\ref{halo-fig} and Table~\ref{halo-tab} show the GCLFs for
the halo clusters in six bandpasses. The Gaussian and $t_5$ distributions
give indistinguishable results for the location of the GCLF peak.
The best-fit Gaussians have smaller values of the dispersion:
the characteristic ${\sigma}_g\approx1.05$ corresponds 
to ${\sigma}_t=0.82$. This is not a serious concern; \citet{sec97}
notes that the theoretical correspondence between ${\sigma}_g$
and ${\sigma}_t$ (${\sigma}_g=1.29{\sigma}_t$) is seldom 
observed for the small number of objects used here.
The $t_5$ form of the GCLF has smaller parameter uncertainties 
than the Gaussian form.
This implies that the $t_5$ function is a better fit to the
GCLFs than the Gaussian, and indeed a comparison of 
the maximum likelihood values shows this to be true in 
almost all cases.
The superiority of the $t_5$ function over the Gaussian
was first pointed out by \citet{sec92}, and confirmed
by SH93. In the remainder of this paper, we consider only the $t_5$
estimates of the GCLF parameters.
The sixth column of Table~\ref{halo-tab}
shows the differences between the mean colors of the objects and the
mean peak colors, measured relative to $V\!$. The differences
are well within the uncertainties of the peak locations, as would be
expected if the halo GCs have no correlation of color with magnitude.
This means that, given sufficiently accurate estimates of the
clusters' mean intrinsic colors (which requires an accurate determination
of the reddening), the GCLF peak in one band can 
be used to estimate peak locations in others bands. 

Next we compare our results to previously published M31 GCLFs, given in
Table~\ref{oldgclf}. We correct the values for the GCLF peak
for our preferred extinction value, $A_V=0.25$. Where the peak
values were given as extinction-corrected absolute magnitudes $M_V^0$, 
we used the corresponding apparent distance moduli ${\mu}_V$ to
convert the values to $V_0^0$. The relation used was
\begin{equation}
V_0^0=M_V^0+{\mu}_V-{A_V}^{\prime}+\Delta A_V
\end{equation} 
where ${A_V}^{\prime}$ is the previous author's
assumed extinction and $\Delta A_V = {A_V}^{\prime}-0.25$. 
[\citet{sl89} estimated extinction internal to M31 for each cluster,
using $R_V=2.65$. We could not correct these extinctions to our values 
because the individual magnitudes and extinctions were not given, but
we did apply a correction for the foreground extinction.]
Our results for the halo GCLF peak and dispersion,
$V^0=16.84\pm0.11$, ${\sigma}_t=0.93\pm0.13$, are consistent 
at the $1\sigma$ level with the results for the $R>10$~kpc sample of \citet{cra85}, 
the halo sample of \citet{sec92}, 
and the `outer' sample of \citet{gne97}. 
We find a brighter peak than \citet{rs79}; their photographic
photometry may be suspect.
Because the number of halo clusters is small, the composition 
of the sample can measurably change the derived GCLF parameters.
If we drop the five `low-probability' clusters mentioned in
Section~\ref{sampdef} from our sample, we recover the
slightly brighter peak value reported by \citet{sec92}.
We are aware of only one measurement of a non-$V$-band GCLF peak in M31:
\citet{st95}  average the $B$ magnitudes of \citeauthor{sec92}'s
halo sample to obtain $<\!\!B\!\!>=17.75\pm0.11$, which corresponds to 
$B_0^0=17.42\pm0.11$. 
This is again consistent with our result at the $1\sigma$ level.

We now turn to the other samples of M31 GCs. Here we consider only
the $V$-band GCLF, as the photometric completeness in other
bands is generally much poorer. We correct the GCLF peaks and dispersions
for the magnitude cutoff at $V=18$. The full and disk samples have similar GCLF peaks
($16.70\pm0.11$, $16.67\pm0.16$) 
and dispersions (${\sigma}_t= 1.11\pm0.09$, $1.18\pm0.12$);
this is unsurprising as the full sample is 77\% disk clusters.\footnote{
The GCLF parameter uncertainties are smaller for the halo than
for the disk sample, even though the disk has about 2.5 times 
as many objects. This is caused by extinction-induced incompleteness: 
when completeness is not corrected for, the uncertainties drop by $\sim50$\%.
In the halo sample, the uncertainties are close to the limit imposed 
by the small numbers: ${\sigma}_g/\sqrt{N}\approx 0.11$. 
}
We find a fainter GCLF peak for the GCS as a whole than do
\citet{cra85} , \citet{sl89}, and \citet{gne97} (considering here the
`all clusters' samples of \citeauthor{cra85} and \citeauthor{gne97}).
There are several likely reasons for this: the older samples are
less complete and have poorer photometry, and their corrections for
incompleteness are either non-existent or do not account for
the extinction effect discussed in Section~\ref{compcorr}.

Neither the full or disk samples has a GCLF peak 
significantly different from that of
the halo, but the disk sample has a $2\sigma$ higher dispersion
than the halo. This is in contrast to previous results:
both \citet{cra85} and \citet{gne97} found 
the GCLF peak to be brighter and the 
GCLF dispersion lower for the inner (`disk') clusters.
We suspected that the peak difference found by other groups
was because they failed to correct for incompleteness due to extinction.
To test this, we tried setting the completeness function to a step
function with the step at $V_0=17.75$. The result was a
much brighter disk peak, resulting in a difference of $0.6$~mag 
between the halo and disk GCLF peaks.
The completeness correction thus has a large effect on
the resulting GCLF, and the extinction cannot be ignored
in a spiral galaxy like M31.
We address the question of lower halo dispersion
in Section~\ref{sec-implic}. 

\begin{figure}
\includegraphics[scale=0.7]{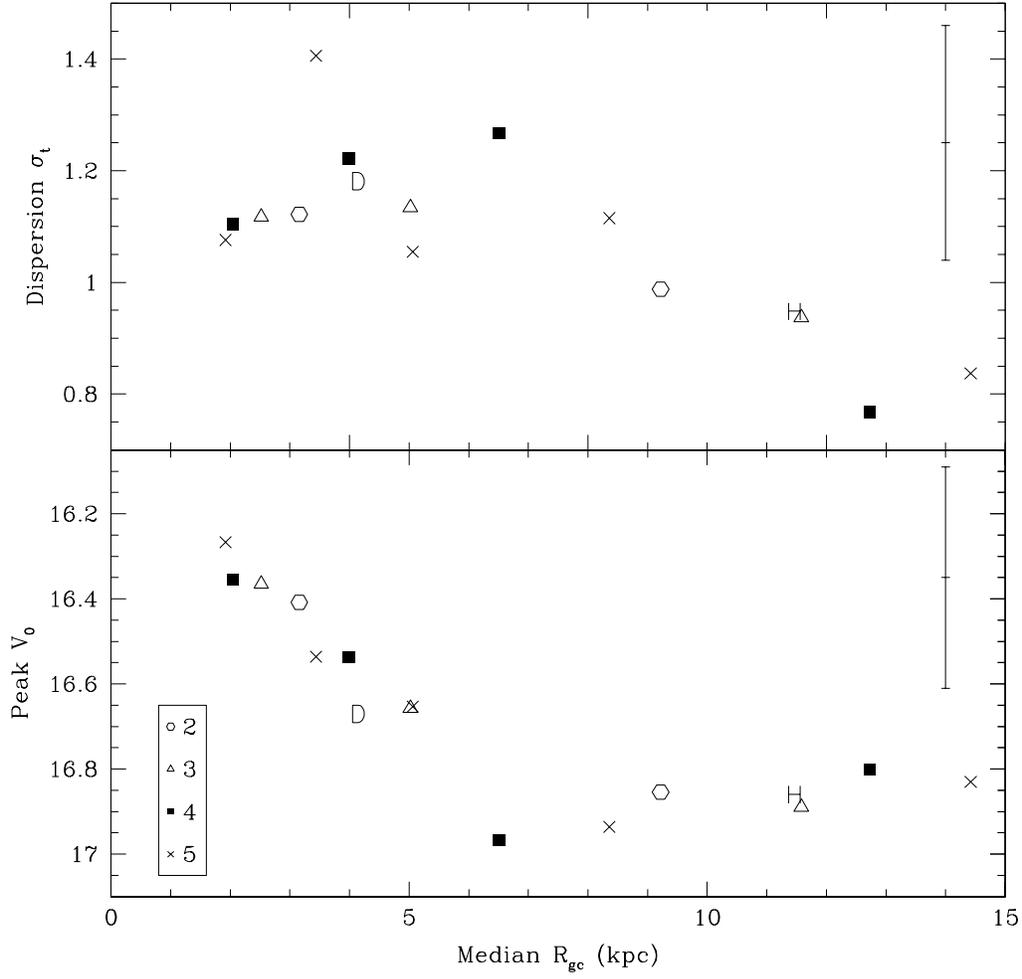}
\caption{GCLF peaks and dispersions for groups of M31 clusters, sorted by projected
distance from the center of M31. The different symbols represent the division of
the GCS into different numbers of groups; `D' and `H' show the location
of the disk and halo samples.\label{rad-fig}}
\end{figure}

Models of GCS evolution predict that low-mass clusters near the 
center of a galaxy are most susceptible to dynamical destruction. 
We suspected that our disk/halo division might be too crude to
detect any radial GCLF differences. 
We followed the method of \citet{gne97} in sorting
the M31 clusters by projected distance from the center of M31
and dividing them into equal-sized groups.
As Figure~\ref{rad-fig} shows, there is a definite
trend for the GCLF peak to be fainter with increasing projected distance,
although the GCLF dispersion does not show any obvious trend.
The difference between the inner- and outer-most groups 
ranges from 0.44 to 0.56 $(\pm0.18)$~mag in $V_0$, depending on the number 
of groups, with $\Delta m^0/{\sigma}(m^0)=2.5-2.8$.
The trend in the peak is largely due to the objects with $R\lesssim3$~kpc.
The disk is dominated by GCs with larger $R_{gc}$, which is why
we see little difference between the disk and halo samples.
This again underscores the importance of having a complete, uncontaminated
sample of the GCs near the M31 nucleus.

\begin{figure}
\includegraphics[scale=0.7]{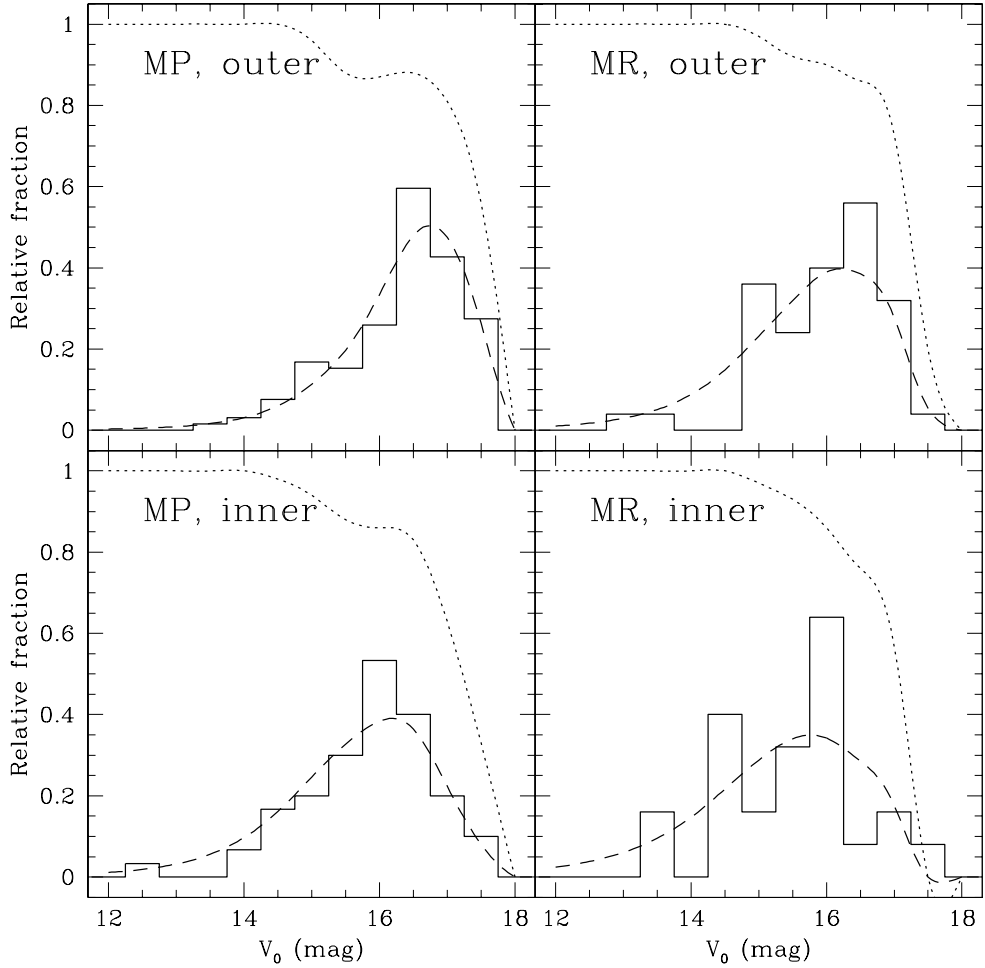}
\caption
{GCLFs and completeness functions for four groups of M31
GCs: (a) MP, outer (b) MR, outer (c) MP, inner (d) MR, inner.
The dip in the completeness function for the MP clusters is
due to the presence of a few bright objects without metallicities
(likely to be metal-poor because MP clusters make up two-thirds
of the population).\label{lf4}}

\end{figure}

Both \citet*{hbk91} and \citet{b00} found that the metal-rich clusters were more
centrally concentrated. Could a difference between MR and MP GCLFs 
produce a radial GCLF difference? To disentangle the effects of age
and metallicity, we divided our full sample by both metallicity and
projected galactocentric distance.
Since the radial GCLF variation is mostly due to the innermost clusters,
we put the inner/outer dividing line at the galactocentric distance containing one-third
of the clusters, $R\approx3.8$~kpc$=19$\arcmin. The resulting GCLF parameters
are in Table~\ref{all-tab}; the GCLFs and completeness functions
are shown in Figure~\ref{lf4}. The significant
difference between GCLF parameters of MR and MP clusters in the
outer two-thirds of M31 strongly implies that both metallicity 
(or some parameter related to it) and galactocentric distance affect the GCLF peak.
To check that the effect we measured was not due to the details of
sample division, we carried out 200 trials in which we divided the 
full sample in both metallicity and $R_{gc}$, with randomly chosen division points.
While the differences between subsamples depended on the exact division point
(as expected), the relative ordering of the GCLF peaks was always the same
as that shown in Table~\ref{all-tab}.

The size of each parameter's contribution to the GCLF peak
can be estimated by writing the values as:
\begin{equation}
\left( \begin{array}{rr} {\rm MP}_i & {\rm MR}_i \\ {\rm MP}_o & {\rm MR}_o \end{array} \right) =
\left( \begin{array}{rr} 16.32 & 16.15 \\ 17.02 &16.46 \end{array} \right) =
\left( \begin{array}{rr} 16.42 & 16.06 \\ 16.92 &16.56 \end{array} \right) +
\left( \begin{array}{rr} -0.10 &  0.10 \\ 0.10  &-0.10 \end{array} \right) 
\end{equation}
so 0.50~mag of GCLF variation is due to galactocentric distance,
0.36~mag to metallicity, and 0.10 to the interaction of the two. 
This is very similar to the average results from the 200 `random division
point' trials.
The existence of the interaction term is not surprising,
since we know that metallicity and galactocentric distance are correlated
\citep[e.g., see Figure~21 of][]{b00}.

\begin{figure}
\includegraphics[scale=0.7]{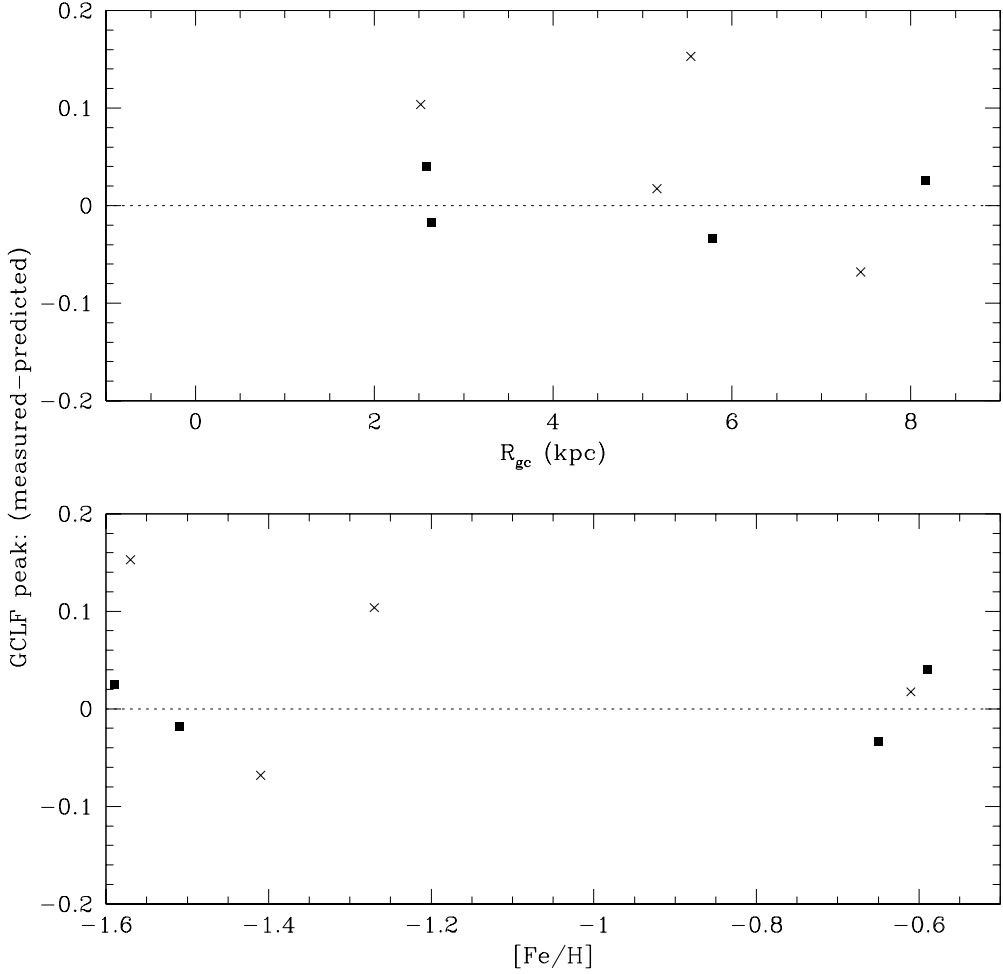}
\caption
{Difference between predicted (from Equation~\ref{eq-reg})
and measured GCLF peaks for the data in Table~\ref{all-tab}. Crosses
are data from the first four lines of the table, used in fitting
the regression equation. Squares are data from the last four lines,
which were not used in the regression because they are not independent
samples of the M31 GC population.\label{regress}}
\end{figure}

Another way to quantify the GCLF variation is a multiple regression of $V^0_0$
on [Fe/H] and projected galactocentric distance $R_{gc}$. 
We performed such a fit for the data in the
first four rows of Table~\ref{all-tab}. Although using four points to define
a plane is not statistically rigorous, it is a reasonable way
to estimate the size of the effects [Fe/H] and $R_{gc}$ have on the GCLF.
(Defining the relation more rigorously will be difficult, since
measuring GCLF parameters with reasonable precision requires sample sizes
$N\gtrsim50$ and the full M31 GC sample of 294 objects can only 
be divided into a few independent subsamples.)
The regression equation is:
\begin{equation}\label{eq-reg}
V^0_0=15.67-0.24{\rm [Fe/H]}+0.12 R_{gc}
\end{equation}
with [Fe/H] in dex and $R_{gc}$ in kpc. This equation predicts $V^0_0$ to within
$\leq 0.15$~mag for the last four samples in Table~\ref{all-tab} (see Figure~\ref{regress}), 
where the clusters are separated on either metallicity or $R_{gc}$ but not both.
A multiple regression of $V_0$ on [Fe/H] and $R_{gc}$ for the individual
cluster data (with no incompleteness correction) gives similar regression parameters, 
but the correlation is much poorer because of the luminosity dispersion. 
This is why we did not did not detect any correlation of luminosity
with metallicity in \citet{b00}, and why we believe that binning the clusters
into samples, although arbitrary, is necessary.

The surprising result here is that the MR clusters are {\em brighter}
than the MP clusters. For clusters of the same age, and average metallicities
${\rm [Fe/H]}\sim -1.6$ and $-0.6$,
\citet{acz95} predict that the MR clusters should be fainter in $V$ by 0.29~mag. 
A difference in age between the two metallicity groups could be responsible, as
younger clusters are expected to be brighter; we return to this point in the
following section. The only systematic error we can conjecture which would cause
overly bright magnitudes for the MR clusters is an overestimation
of their extinction. However, this would also make the derived
intrinsic colors of the MR clusters too blue, which seems unlikely 
given that the color-metallicity
relations for Galactic and M31 clusters match well throughout
their metallicity range \citep[see][]{b00}.

Reports of large GCLF variations with either $R_{gc}$ or metallicity
are not common in the literature. 
\citet{arm89} assigned Milky Way GCs to the
disk or halo based on [Fe/H] (with a dividing line at $-0.80$~dex) 
and found no difference in the luminosity functions of the two groups.
\citet{kav97} found no difference in GCLF peak magnitudes
of the inner and outer halo clusters in the Milky Way and M31; \citet{gne97} 
found a 0.47~mag difference between GCLF peaks of the same samples and claimed
that the histogram-fitting method used by \citeauthor{kav97} was responsible
for their failure to detect the difference.
\citet{gne97} also found a difference of $0.79\pm0.12$~mag 
in GCLF peak magnitudes between inner and outer GCs in M31. 
This difference is larger than the one we measure, but certainly
compatible with it at the $1\sigma$ level.
Selection effects in our M31 GC catalog are a possible
cause of the differences between our results and those of previous
authors; we consider this in the following section.

Most published GCLFs are actually for elliptical galaxies --
M31 and the Milky Way are the only two spirals whose GCLFs
are well-measured (but these two galaxies are often used, 
e.g. by \citet{st95}, to define the GCLF peak absolute magnitude
for use in distance measurement).
In ellipticals where GCLFs have been determined
separately for the red (MR) and blue (MP) GCs \citep[e.g.][]{kws99,puz99,gri99}
the metal-poor clusters are typically found to be brighter (0.1--0.5~mag)
in $V$, and fainter or similar in $I$, as predicted by 
\citet{acz95}. \citet{lk00} analyzed the same HST data as \citet{puz99}, 
but found that GCLF peaks of the red and blue clusters were indistinguishable
in both $V$ and $I$.
\citet{gne97} found a $0.24\pm0.11$~mag
difference between the GCLF peaks of inner and outer clusters in M87, while
\citet{kws99} found no evidence for radial variation in the M87 GCLF 
and suggested that \citeauthor{gne97} did not correctly account for
incompleteness. \citet{puz99} found that only the red clusters 
in NGC~4472 show a difference between inner and outer GCLF peaks
($0.27\pm0.14$ in $V$). Our results on the radial GCLF variation are
not inconsistent with those of other authors, but the
variation with metallicity or color has not been reported before.
It is important to know whether the GCLF variation we measure
is a unique feature of the M31 GCS, a common feature of spiral
galaxies' GCSs not shared by the Milky Way, or an artifact of
the data and methods we used. A definitive answer to this 
question will require better data on the M31 clusters and
their individual extinctions and information on the GCLF
in other spirals.

\subsection{Selection effects and the GCLF\label{contam}}

Because M31 GC catalogs, including ours, suffer from
both incompleteness and contamination, it is reasonable to ask how 
the catalog properties might affect the GCLF. 
Specifically, could variations in the degree of contamination and incompleteness
with galactocentric distance be responsible for the radial GCLF variations
we measured?

\begin{figure}
\includegraphics[scale=0.7]{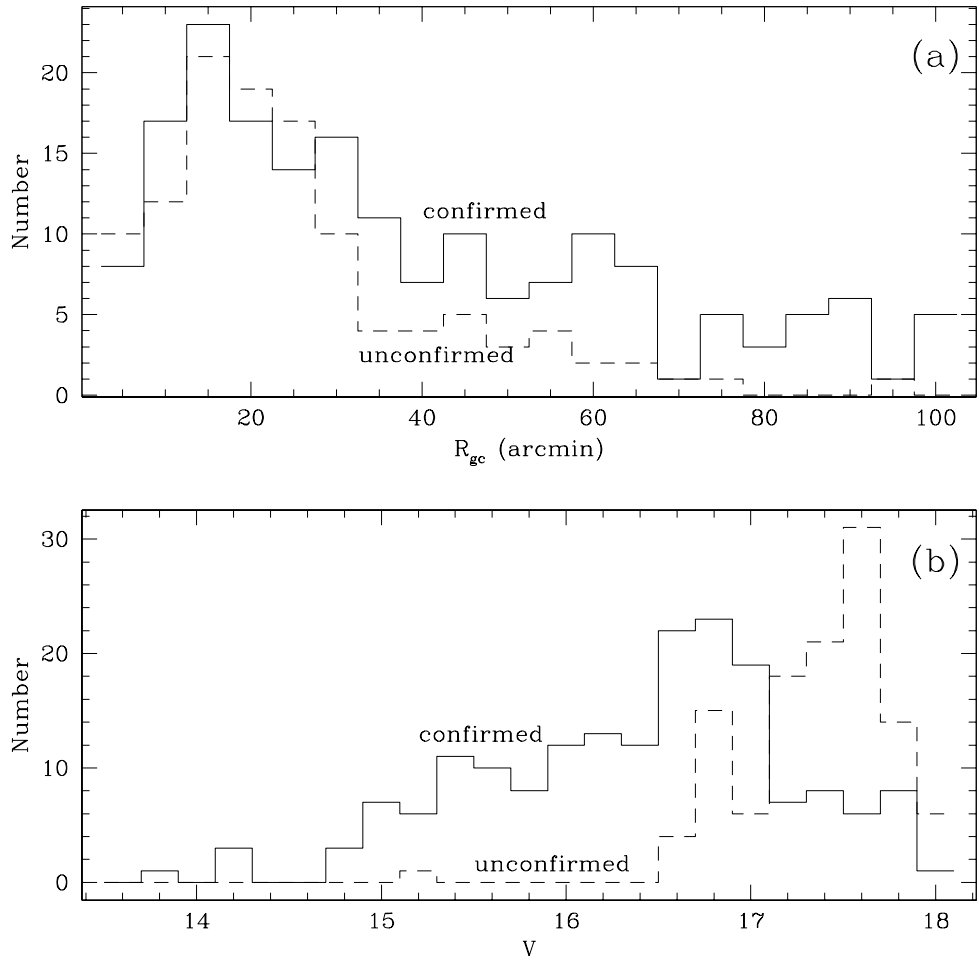}
\caption
{Distribution of $R_{gc}$ and $V$ for confirmed and unconfirmed
M31 GCs. The fraction of confirmed clusters rises slowly with
$R_{gc}$ and drops rapidly with $V$.\label{conf}}
\end{figure}

A small fraction of the objects in our cluster catalog are probably not M31 GCs. 
Of the 294 objects
in our full sample, 178 (about 60\%) have been confirmed as clusters by
spectroscopy or high-resolution imaging. Most of the halo sample (73/86)
is confirmed, and Figure~\ref{conf}a shows that the fraction of confirmed
clusters is highest in the outer regions. The confirmed objects also tend
to be brighter, as Figure~\ref{conf}b shows: this is not surprising
since the largest spectroscopic survey \citep[that of][]{hbk91} was performed
descending a magnitude-ordered list. We are not aware of any `false positives' 
--- objects whose confirmation as a GC was later shown to be incorrect --- 
among M31 GC candidates, so any non-clusters must therefore be 
among the 116 unconfirmed objects. Since these objects are, on average, fainter
than the confirmed clusters, dropping contaminating objects from the sample
should make the resulting GCLFs brighter. A larger effect would be expected for
the inner region, since it has more unconfirmed objects.

We tested this hypothesis using a Monte Carlo experiment.
For the full sample and each of the subsamples listed in Table~\ref{all-tab},
we performed 200 trials in which we chose 34\% of the unconfirmed clusters at random, 
dropped them from the sample, and computed the GCLF with the remaining objects. 
The 34\% figure comes from the \citet{bat87} classification of these objects and 
\citeauthor{rac91}'s (\citeyear{rac91}) estimate of the actual cluster 
fractions in each class. Table~\ref{mc-tab} shows the average of the GCLF
parameters over all 200 trials. Compared to the full samples (Table~\ref{all-tab}), 
the GCLFs found for the `decontaminated' samples had brighter average peaks and 
almost identical dispersions; this was expected because most of the 
unconfirmed clusters are faint. The change caused by decontamination was was 
larger for the inner and metal-rich clusters; again, this was expected
because these clusters are less likely to be confirmed.
Table~\ref{diff-tab} shows that decontaminating the cluster catalog 
actually increases the GCLF peak and dispersion differences 
between different samples.
We conclude that contamination of the GCLF sample by non-clusters cannot be responsible
for the GCLF differences we measure.

A first attempt at estimating the catalog incompleteness can be made
using the work of \citet{m98}.  They searched for
globular clusters in four fields on the eastern side of the M31
disk ($R_{gc}\sim$20--40\arcmin ) with a total area of 440.7 
square arcmin. \citeauthor{m98} discovered 67 new GC candidates in this region, 
but considered all except five to be low-probability candidates. Two of 
the five good candidates had $V<18$ and three had $V>18.5$. 
The area of the the M31 disk
(the large ellipse in Figure~\ref{magcomp-ir}) is $1.14\times 10^4$ square
arcmin, so can we scale the \citet{m98} result by area to
estimate that our catalog could be missing as many as 50 clusters 
with $V<18$. We do not have any information on the magnitude distribution
of the possible missing clusters, other than the fact that
virtually all of \citeauthor{m98}'s new cluster candidates have $V\gtrsim17.5$.
If our catalog were truly missing 50 objects, it would be about 83\% 
complete. Because this value is a large extrapolation from very 
few data points, it can only be considered a rough estimate of
our catalog's completeness, and it yields no information on 
completeness variation with magnitude or galactocentric distance.

An incompleteness estimate based on the work of \citet{m98}
is suitable for the outskirts of M31, but not
for the crowded inner region. To estimate the number of missing GCs 
and contaminating objects in this regions, we used data in the
HST Archive to search for new and previously known M31 GCs; 
the details of this effort will be described
in a subsequent paper (Barmby \& Huchra, in preparation). 
Briefly, we examined all 79 WFPC2 fields with $R_{gc}\leq30$\arcmin\ 
imaged with broadband optical filters (F300W to F814W) 
and with exposure times $\geq100$~s. We chose $R_{gc}=30$\arcmin\
as the outer limit because the fractional area 
of M31 covered by HST fields is only 5\% at this distance and the
chances of finding globular clusters in more distant fields are small.
Working independently, two of us (PB and JH) 
visually searched each image for globular clusters. 
The searches were conducted `blind', i.e., with no reference to the positions
of known M31 GCs.  
To estimate our cluster detection efficiency, we inserted `artificial' globular 
clusters, made by randomly rotating and re-scaling HST images of
known M31 globulars, into some of the images.
Our overall success rate for detection of the artificial
clusters over the magnitude range $14<V<22$ was 80\%; 
for objects with $V<18$, our primary concern for this work,
our success rate was 96\%. We conclude that we should have detected
essentially all true M31 GCs with $V<18$ in the HST images.

\begin{figure}
\includegraphics[scale=0.7]{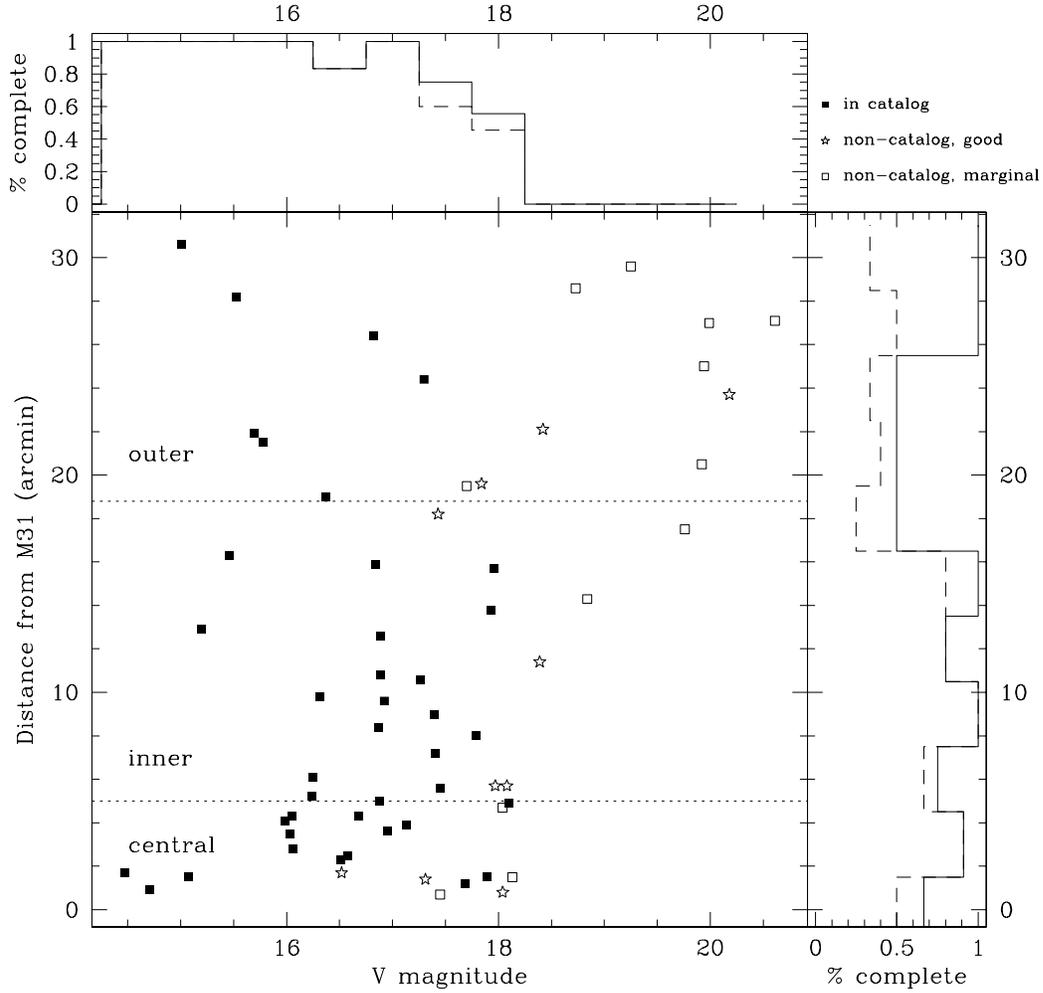}
\caption{Distribution in $R_{gc}$ and $V$ for cluster candidates
in HST fields. The catalog completeness as a function of
$R_{gc}$ and $V$ is (number of cataloged clusters) divided by
(total number of clusters); solid lines are the completeness
computed using good GC candidates only and dashed lines include
marginal candidates.\label{fig-hst}}
\end{figure}

Our final list of cluster candidates contains 62 objects. 40 of these are 
previously-discovered objects appearing in our catalog \citep{b00} 
of plausible cluster and candidates, 9 are previously-discovered objects
not in our catalog because they had been classified as `low-probability' candidates,
and 13 objects are newly-discovered cluster candidates. 
We estimated $V$ magnitudes for the new candidates from the HST images, using
the transformations given in \citet{h95}. 
Figure~\ref{fig-hst} shows the distribution of the HST cluster candidates in $V$ 
and $R_{gc}$. Only about half of the non-cataloged objects are unquestionably
globular clusters, so we mark `good' and `marginal' candidates with
different symbols.
Most of the newly-discovered candidates are either faint objects at 
large galactocentric distances, or bright objects very close to the galaxy center.
The discovery of bright new cluster candidates having $R_{gc}\leq5$\arcmin\ 
shows that incompleteness is substantial in the region, and validates
our decision to exclude it from GCLF computations.

The important question here is whether there is a difference
in catalog completeness between inner and outer clusters with $V<18$. 
In the inner region there were 20 
good GC candidates with $V<18$; 17 of these objects were in our catalog, so its 
completeness is 85\%. 
In the outer region there were 8 good GC candidates 
with $V<18$; only one of these objects was not in our catalog, so the
completeness is 88\%. Although the numbers are small because the covering 
fraction of the HST images on the sky is only 10--20\%, there is no evidence that the
inner sample of clusters is substantially less complete than the outer one.
The new cluster candidates in both regions have similar magnitudes,
$V\sim17.7$, indicating that the magnitude limit of the existing 
catalogs does not change drastically with $R_{gc}$.

To the best of our ability to model
them, the selection effects of incompleteness and contamination in 
the M31 cluster sample are not responsible for the GCLF variations which 
we measure. The M31 GC catalog can certainly be improved,
and our conclusions would be strengthened if the
same GCLF differences were measured in less-contaminated samples with
better-understood and spatially uniform completeness.
However, analysis of the GCLF does not require a perfect catalog ---
something which does not exist even for the Milky Way GCs.
As a final comment on catalog incompleteness, we speculate on the possibility
that there exists a heavily-obscured population of GCs in M31.
If such objects exist and are fainter than the known population
{\em and\/} preferentially located near the center of M31 or metal-rich,
then the brighter peaks found for the metal-rich and inner clusters might 
be an artifact of incompleteness. There is presently no evidence for a population
of heavily-extinguished GCs in M31, so we believe our GCLF
results to be valid, and consider their implications in the following section.

\subsection{Implications for GCS destruction and formation models\label{sec-implic}}

Variation of the GCLF with $R_{gc}$ is a key prediction of
GCS evolution models.
The shorter destruction timescales for clusters near the center of
a galaxy potential may lead to a mass difference, and hence a GCLF
difference, between `inner' and `outer' GCs in a galaxy.
Several authors have predicted differences between the 
inner and outer GCLFs for Milky Way clusters. 
The dividing line between inner and outer varies from 5--10~kpc,
and the size of the predicted difference in the GCLF peaks ranges
from essentially zero \citep{v98} to 1.4 magnitudes \citep{b97}.
\citet{v98} finds that there is a particular initial GCMF for which
the balance between cluster disruption and mass evolution of 
the surviving clusters keeps the shape and initial parameters of
the mass function unchanged.
The only prediction made specifically for the M31 GCS is that of
\citet{og97}. For an inner/outer split at $R=27\arcmin = 5.5$~kpc,\footnote{
It is unclear whether this value is a projected or true 3-D 
distance from the center of M31; we assume it to be a projected distance.} 
they predict a peak location difference of $0.49\pm0.18$.
Our results on the GCLF peak are in in excellent agreement
with this prediction.
For two populations of the same age and metallicity, a magnitude 
difference of 0.5 corresponds 
to a mass ratio of $10^{0.2}=1.58$. For populations with
the same age and metallicities
differing by 1~dex, a magnitude difference of 0.35 in
$V$ implies a mass ratio of 1.87; the larger mass ratio is due to
the increase of mass-to-light ratio with metallicity.

\citet{og97} predict that destruction of low-mass inner clusters
results in a (Gaussian) GCLF dispersion for the inner clusters 
$0.17\pm0.12$~mag  lower than that of the outer clusters.
\citet{v98} also predicts a lower inner GCLF
dispersion for most initial conditions. We do not measure a lower 
GCLF dispersion for the inner clusters: if anything, 
the dispersion seems to decrease, not increase, with projected distance.
While photometric and extinction errors are probably larger
for the inner clusters, we showed in Section~\ref{compcorr} that
only very large errors strongly bias the GCLF parameters. The completeness limit
is brighter for the inner clusters; again, this should not affect the
GCLF parameters since we correct for the resulting bias.
We are unable to devise any other systematic effects which might
lead to erroneous measurements of the GCLF dispersion. 
A larger age spread in the inner clusters might mask a decrease
in GCLF dispersion. If such an age spread existed in M31, it would be in
contrast to the Galactic GCS, where the age spread is larger
for the outer clusters \citep*{cds96}. It is also possible that
the inner clusters had a log-normal initial GCMF with parameters
that resulted in evolution in mean mass but not in dispersion
\citep[see][]{v98}.
Since the theoretically predicted dispersion differences are
small compared to the peak location differences, we do not consider
our failure to detect such differences a serious problem for
either our method or the theoretical models.

Key assumptions made when calculating the effects of 
dynamical destruction on the GCLF are that the initial
age and mass distributions of the GCs are not functions 
of $R_{gc}$, and that metallicity effects are not important. 
These assumptions may not be correct.
\citet{bh00} found evidence that the metal-rich 
clusters in M31 and the Milky Way were younger than the  
metal-poor clusters; \citet{ros99} found similar results 
from the color-magnitude diagrams for a sample of Milky Way
clusters. If age differences do exist, they should affect the GCLF.
To estimate the size of the age difference, we examined the 
predictions of $V$-band luminosity in the models of \citet{bc96},
\citet*{kff99}, and \citet{w96}. We computed the age differences 
implied by $\Delta V =0.35$~mag, with (fainter, brighter) populations at 
${\rm [Fe/H]} = (-1.63,-0.63)$ \citep{bc96,kff99} or $(-1.50,-0.50)$ \citep{w96}.
\citeauthor{w96}'s models cover only the age range from 8--16~Gyr 
at these low metallicities, and the only age pair with the required $\Delta V$
is ${\rm {age}_{MR}}=8$~Gyr, ${\rm {age}_{MP}}=16$~Gyr. For the other
two sets of models, which cover the age range 1--16~Gyr, 
several age pairs reproduce the GCLF peak difference. 
In each case the MR clusters are about half as old as the MP clusters.
Explaining the GCLF difference of outer and inner clusters
as an age difference is also possible. For populations of the
same metallicity (the median [Fe/H] values of the inner and outer clusters
are not very different at $-1.27$ and $-1.41$), a 0.5~mag 
difference in $V$ corresponds to the brighter clusters being 55\%
as old as the fainter clusters.

If the two sets of clusters have the same stellar initial mass function 
(IMF), the effect of IMF on the derived age differences is fairly small.
For a \citet{sca86} or \citet{ms79} IMF instead of \citet{sal55}, 
$\Delta V=0.35$ at different metallicities makes the MR clusters 
about 55\% as old as the MP clusters,
and $\Delta V=0.5$ at the same metallicity makes the bright clusters
60\% as old as the fainter ones. 
However, if the two sets of clusters had {\em different} IMFs,
this could produce a large difference in the average
$V$ magnitudes. All three models predict that Scalo-IMF
populations should be brighter than Salpeter-IMF populations of the 
same age and mass, by $0.6-1.0$~mag in $V$ for ages $>8$~Gyr. A difference in 
IMF with metallicity would still require explanation, however;
at least for Galactic GCs, \citet{dp99} find no evidence of 
IMF variation.

Independent constraints on cluster ages and IMFs are needed
in order for GCLF variations to be used in the study of
GCMF evolution. If different populations start with the same GCMF, 
differing dynamical evolution could lead to mass differences
and explain the observed GCLF differences.
However, combinations of age, metallicity
and IMF differences could also reproduce the observations. 
With only one observable it is not possible to constrain all 
of the cluster parameters. Measuring the GCLF variations
in the $K$-band, which is less sensitive to age and 
more sensitive to metallicity \citep[the mass-to-light ratio 
in $K$ is expected to decrease with metallicity, rather than increase
as it does in $V$;][]{w94} would be helpful in disentangling the two effects. 
Unfortunately, our catalog lacks near-IR photometry for many clusters in
the disk, so it is not possible to accurately 
measure the near-IR GCLF for the inner clusters.

Examining the measured GCLF parameters leads to several important conclusions.
The fact that metal-rich clusters are, on average, brighter than metal-poor
clusters implies that there must be some mass, age, or IMF difference between
the two groups. We consider here the extreme possibilities
for age and mass differences that could reproduce this effect
(we neglect possible IMF differences, since these seem the least likely).
Either (1) the MR clusters are younger by $\sim50$\%, or
(2) the MR clusters are more massive by a factor of $\sim1.9$.
The fact that inner MP clusters are brighter than outer MP clusters
means that there is also a GCLF variation with $R_{gc}$.
Again, the two extreme possibilities are:
(3) the inner clusters are younger by $\sim45$\%, or
(4) the inner clusters are more massive by a factor of $\sim1.6$.
Each of these possibilities has important implications for 
the understanding of GCS and galaxy formation and evolution; we examine 
each in turn, drawing on the review given by \citet{kp00}.

1) If the MR clusters are younger than the MP clusters, 
an age-metallicity relation exists in the M31 GCS. Such a
relation has long been suspected for the Galactic clusters,
although it has not been demonstrated conclusively \citep*[see][]{scd97}.
\citet{lk00} and \citet{kws99} compared their GCLF measurements in 
NGC~4472 and M87, respectively, to population synthesis models.
Both found that, in addition to the metallicity effect,
younger ages for the metal-rich clusters 
were required to account for the observed GCLF peak differences. 
A relation between age and metallicity is consistent with several 
scenarios for the formation of globular cluster systems. These include
the merger scenario of \citet{az92}, the in situ/two-phase scenario of
\citet*{fbg97}, and the pre-galactic scenario of \citet{kp97} \citep[see also][]{pd68}. 
Fitting young MR GCs in M31 into the merger scenario would
require some modifications. The scenario attempts
to account for the presence of multiple populations of GCs in
{\em elliptical} galaxies by postulating that the MR clusters form 
when star formation is induced in
the spiral/spiral merger which produces the elliptical.
M31 would thus have to be a merger product itself. While there 
is little evidence that M31 has had a recent major merger, 
perhaps the `minor merger' 
of a satellite galaxy with M31 could have excited star formation
in the M31 disk \citep{hm95}, generating a population of
younger GCs.

\citet*{fbg97} suggested that the two populations of GCs in 
ellipticals formed in two distinct phases 
separated by several Gyr, with the disk GCs in spirals
representing a third collapse phase. The pre-galactic scenario also
has two phases, but the difference is that the
MP clusters are unrelated to the final galaxy.
Either scenario is is compatible with our results, although
neither specifically predicts the large age difference we estimate 
or the relative number of MR and MP clusters.
It is interesting that all three scenarios were devised to explain the multiple
GC populations of ellipticals, but there is, so far, little evidence
for any age-metallicity relation in the GCs of such galaxies.
\citet*{cbr98} found that their samples of MR and MP GCs in M87
are coeval and old (although they suggested that there may be a problem with
their model calibration). \citet{puz99} and \citet{kp98} found similar results
for NGC~4472 and NGC~1399, respectively.

2) Is the mass of M31 GCs somehow related to their metallicity?
Massive clusters could have their metallicity increased by self-enrichment,
and self-enriched GCs must be massive enough
to survive disruption by supernova explosions.
\citet{ml89} give a minimum mass for surviving GCs of $10^{4.6}$~M$_{\sun}$, 
low enough to include almost all Galactic and M31 GCs. They state
that ``it is difficult to `predict' any trend of metallicity with mass
without extra information on the IMF'', but it seems logical that 
more massive clusters would generate a larger first generation of
stars and hence have more enrichment. If self-enrichment occurred
in Galactic GCs, it had to be rapid enough to produce the
chemical homogeneity observed in most clusters \citep[e.g.][]{sun93}.
\citet{p99} claim that, contrary to previous conclusions,
the formation time for the second generation of stars is longer
than the mixing timescale; if true, this means that self-enrichment 
(and a mass-metallicity relation) a realistic possibility.
Instead of larger masses causing higher metallicity, could higher-metallicity 
gas induce the formation of more massive GCs?
If metallicity is important in GC formation, the opposite
seems more likely to be true. In models relying on a 
cooling condition \citep[e.g.][]{ml92}, MR clusters 
are predicted to be less massive (for the same IMF) 
because cooling is more efficient in high-metallicity clouds. 
An age-metallicity relation with a small mass-metallicity
contribution from self-enrichment is our preferred scenario
for explaining the GCLF differences between MR and MP GCs in M31.

3) The possible explanations for younger inner clusters are
similar to those for younger metal-rich clusters. 
The inner clusters could `naturally' have formed much later than the
outer clusters (during the end of the galaxy collapse phase?),
or perhaps some external event caused the formation
of inner GCs much later than the bulk of the population.
Another possibility is that the inner clusters were accreted
along with dwarf galaxies cannibalized by M31, although
there is no obvious reason why dwarf galaxy GCs should be younger
than those in larger galaxies. All of these scenarios are rather 
{\em ad hoc} and none are obviously related to theoretical ideas for 
globular cluster system formation. We do not consider
any to be well-supported.

4) Globular cluster mass and galactocentric distance might
be causally connected in either direction.
Low-mass clusters are thought to be more easily destroyed near the 
galaxy center, leading to a larger average mass. This is 
is one of the major predictions of models of GC destruction.
Although model predictions vary widely, at least some models
\citep[e.g.][]{og97} predict values for the GCLF difference
close to what we observe. We do not measure the
model-predicted difference in GCLF dispersion, but
this may not be a serious problem.
Could conditions near the galaxy center affect the
average GC mass at formation? 
In the \citet{ml92} GC formation scenario, GC mass is expected to
{\em increase} with galactocentric distance, and we are aware of no
models which explicitly predict a decrease in mean GC mass with distance.
GC destruction by dynamical effects in the inner part of M31 
is our preferred scenario for explaining the GCLF differences 
between inner and outer GCs in M31.

\section{Conclusions}

We have calculated the first $URJK$ GCLFs for M31 halo globular clusters,
and find that the GCLF peak colors are consistent with the average 
cluster colors. Our parameters for the $V$- and $B$-band halo GCLFs are consistent 
with those of other groups. We find no significant differences between the 
disk and halo GCLF peaks, although the disk has a lower dispersion.
A difference in GCLF peak at the $2\sigma$ level occurs when we consider 
the inner and outer-most groups, as determined by projected galactocentric distance. 
This difference is consistent with that predicted by \citet{og97} for M31;
however, we do not detect the predicted difference in GCLF dispersion.
We separate the M31 clusters by metallicity and find that the metal-rich
clusters have a brighter GCLF peak than the metal-poor clusters, even
when the difference in $R_{gc}$ is taken into account.
Modeling of the catalog selection effects suggests that these
effects are not responsible for the measured GCLF differences.
However, an M31 GC catalog with well-understood and spatially uniform 
completeness and contamination is required
in order to definitively confirm our results. Such a catalog might be produced by 
a near-IR, high spatial resolution survey of M31.
We consider the implications of the GCLF differences for
models of globular cluster and GCS formation, and conclude that
younger ages for metal-rich clusters plus dynamical destruction of
inner clusters are the most likely causes of the observed GCLF variations.

\acknowledgments

We thank the referee for incisive comments,  
G. Smith and L. Schroder for help with the Lick observations, and
J. Secker for making his maximum-likelihood GCLF code available.
We thank the staffs of Whipple and Lick Observatories for productive 
observing runs and patient post-observing support.
This research was supported by the Smithsonian
Institution, Faculty Research Funds from the University of California, 
Santa Cruz, and NSF grant AST-990732.

\clearpage

\clearpage

\begin{deluxetable}{lllll}
\tablewidth{0pt}
\tablenum{1}
\tablecaption{New infrared photometry for M31 GCs\label{newphot}}
\tablehead{\colhead{name}&\colhead{$J$}&\colhead{$H$}&\colhead{$K$}}
\startdata
005-052 &   14.14(1) & 13.78(2) & 13.54(3) \\
150D    &   15.72(1) & 15.17(1) & 15.16(3)  \\ 
167D    &   16.33(2) & 15.69(2) & 15.57(6)  \\ 
196-246 &   15.37(1) & 14.73(2) & 14.60(3)  \\ 
208-259 &   15.39(4) & 14.71(4) & 14.54(6)  \\ 
232-286 &   14.05(1) & 13.52(1) & 13.34(2)  \\ 
236-298 &   15.67(2) & 14.99(2) & 15.05(5)  \\ 
242D    &   16.47(3) & 16.32(5) & 16.09(8)  \\ 
316-040 &   15.36(4) &  \nodata & \nodata    \\ 
320-000 &   15.44(2) &  \nodata & 14.15(3)  \\ 
328-054 &   15.82(9) &  \nodata & \nodata    \\ 
329-000 &   15.94(3) &  \nodata & 14.84(6)  \\ 
330-056 &   15.84(7) &  \nodata & \nodata    \\ 
331-057 &   15.92(9) &  \nodata & 14.89(12) \\  
333-000 &   16.36(10)&  \nodata & \nodata    \\ 
336-067 &   16.14(2) & 15.78(3) & 15.65(7)  \\ 
344D    &   15.39(7) &  \nodata & \nodata    \\ 
384-319 &   13.87(1) & 13.14(1) & 12.91(1)  \\ 
399-342 &   15.69(2) & 15.14(2) & 15.07(4)  \\ 
423-000 &   16.10(2) & 15.48(2) & 15.66(3)  \\ 
450-000 &   16.65(5) &  \nodata  & 15.69(11) \\ 
457-097 &   15.38(2) & 14.79(2) & 14.79(3)  \\ 
461-131 &   15.43(1) & 14.78(1) & 14.68(3)  \\ 
469-220 &   15.92(2) & 15.42(3) & 14.92(4)  \\ 
472-D064&   13.46(1) & 12.84(1) & 12.63(2)  \\ 
BA11    &   15.98(2) & 15.65(3) & 15.19(4)  \\ 
DAO054  &   15.94(2) &  \nodata  & 14.90(4)  \\ 
DAO055  &   17.02(7) &  \nodata & \nodata    \\ 
DAO094  &   15.88(3) &  \nodata & 14.84(5)  \\ 
DAO104  &   16.86(5) &  \nodata & 15.82(11) \\ 
\sidehead{Objects with previous photometry}
024-082 &   14.77(1) & 14.13(1) & 13.99(1)  \\ 
027-087 &   13.85(1) & 13.20(2) & 13.13(2)  \\ 
042-104 &   13.21(1) & 12.44(1) & 12.28(1)  \\ 
068-130 &   13.67(1) & 12.86(1) & 12.67(1)  \\ 
082-144 &   12.43(1) & 11.73(1) & 11.44(1)  \\ 
217-269 &   14.47(2) & 13.82(2) & 13.66(3)  \\ 
298-021 &   14.97(1) & 14.44(1) & 14.39(3)  \\ 
317-041 &   15.05(1) & 14.62(2) & 14.54(3)  \\ 
\enddata
\tablecomments{Objects with $H$-band measurements were observed
at Lick; others were observed at FLWO. Bracketed numbers are measurement
uncertainties in hundredths of a magnitude.}
\end{deluxetable}

\begin{deluxetable}{lllllll}
\tablewidth{0pt}
\tablenum{2}
\tablecaption{M31 Halo GCLF\label{halo-tab}}
\tablehead{\colhead{bandpass}&\colhead{$m^0_g$}&\colhead{${\sigma}_g$}&\colhead{$m^0_t$}&\colhead{${\sigma}_t$}&\colhead{${\Delta}_t$}&\colhead{$N$}}
\startdata
$U_0$ & $17.68\pm0.21$	& $1.05\pm0.15$	 & $17.65\pm0.13$  & $0.87\pm0.14$ &0.01   &74 \\
$B_0$ & $17.61\pm0.21$	& $1.12\pm0.17$	 & $17.57\pm0.14$  & $0.95\pm0.16$ &0.07   &80 \\
$V_0$ & $16.81\pm0.13$	& $1.06\pm0.10$	 & $16.84\pm0.11$  & $0.93\pm0.13$ &\nodata&86 \\
$R_0$ & $16.42\pm0.19$	& $1.10\pm0.14$	 & $16.40\pm0.14$  & $0.95\pm0.14$ &0.05   &68 \\
$J_0$ & $15.26\pm0.20$	& $1.08\pm0.14$	 & $15.26\pm0.14$  & $0.98\pm0.16$ &0.09 &73 \\
$K_0$ & $14.39\pm0.20$	& $0.97\pm0.15$	 & $14.45\pm0.18$  & $0.89\pm0.15$ &0.04 &67 \\
\enddata
\tablecomments{$\Delta=\overline{(V-X)_0}-(V_0^0-X_0^0)$ is the difference between the mean object
color and the peak color. The superscript in $V^0_0$ refers to the GCLF peak,
while the subscript refers to the extinction correction \citep{har88}.}
\end{deluxetable}

\begin{deluxetable}{lllclc}
\tablewidth{0pt}
\tablecaption{M31 $V$-band GCLF parameters\label{oldgclf}}
\tablenum{3}
\tablehead{\colhead{sample}&\colhead{$V^0$}&\colhead{$\sigma$ (Gaussian)}
&\colhead{$N$\tablenotemark{a}}&Ext. correction\tablenotemark{b}&\colhead{Ref.}} 
\startdata

halo &$16.54\pm0.12$&1.13&86&fg&(1)\\
all  &16.95&1.2&188&fg&(2)\\
all       &16.5&1.2&408&\tablenotemark{c}&(3)\\
$R<10$~kpc&16.2&1.5&265&\tablenotemark{c}&(3)\\
$R>10$~kpc&17.0&1.8&143&\tablenotemark{c}&(3)\\
all  &$16.13\pm0.08$&$1.16\pm0.08$&294&M31+fg&(4)\\
halo &$16.75\pm0.11$&$1.10\pm0.11$&82&fg&(5)\\ 
halo &$16.75\pm0.15$&$1.10\pm0.11$&82&fg&(6)\\
halo &$16.75\pm0.15$& \nodata     &81&fg&(7)\\
      halo&$17.20\pm0.08$&$0.76\pm0.05$&161&fg&(8)\\
inner halo&$17.05\pm0.10$&$0.57\pm0.10$&64 &fg&(8)\\
outer halo&$17.04\pm0.17$&$0.94\pm0.13$&97 &fg&(8)\\
all  &$16.27\pm0.06$&$0.86\pm0.04$&164&\tablenotemark{c} &(9)\\
inner&$15.92\pm0.08$&$0.70\pm0.05$ &82&\tablenotemark{c} &(9)\\
outer&$16.71\pm0.09$&$0.96\pm0.08$ &82&\tablenotemark{c} &(9)\\
\enddata
\tablenotetext{a}{$N$ is the number of objects used in the GCLF measurement, not
an estimate of the total number of GCs $N_{gc}$.}
\tablenotetext{b}{`fg' indicates correction for foreground
extinction of $A_V=0.25$~mag} 
\tablenotetext{c}{Paper text implies that these values are
corrected for extinction, but method not clearly described.}
\tablerefs{(1) \citet{rs79}; (2) \citet{svdb85}; (3) \citet{cra85};
(4) \citet{sl89}; (5) \citet{sec92}; (6) \citet{rac92}; (7) \citet{rhh94}; 
(8) \citet{kav97}; (9) \citet{gne97}}
\end{deluxetable}

\begin{deluxetable}{llllll}
\tablewidth{0pt}
\tablenum{4}
\tablecaption{GCLF for different samples of M31 GCs\label{all-tab}}
\tablehead{\colhead{Sample}&\colhead{$m^0_t$}&\colhead{${\sigma}_t$}&\colhead{[Fe/H]}
&\colhead{$R_{gc}$ (kpc)}&\colhead{$N$}}
\startdata
MP inner &$16.32\pm0.21$ &$1.00\pm0.17$&$-1.51$& 2.6&60 \\
MP outer &$17.02\pm0.22$ &$1.08\pm0.13$&$-1.59$& 8.2&131\\
MR inner &$16.15\pm0.58$ &$1.51\pm0.32$&$-0.59$& 2.6&25 \\
MR outer &$16.46\pm0.23$ &$0.93\pm0.18$&$-0.65$& 5.8&50 \\
MP       &$16.84\pm0.16$ &$1.09\pm0.11$&$-1.57$& 5.5&191\\
MR       &$16.43\pm0.27$ &$1.11\pm0.19$&$-0.61$& 5.2&75 \\
inner    &$16.37\pm0.21$ &$1.12\pm0.17$&$-1.27$& 2.5&98 \\
outer    &$16.80\pm0.14$ &$1.05\pm0.10$&$-1.41$& 7.4&196\\
\enddata
\tablecomments{[Fe/H] and $R_{gc}$ are median values.}
\end{deluxetable}

\begin{deluxetable}{lll}
\tablewidth{0pt}
\tablenum{5}
\tablecaption{Average GCLF parameters for de-contaminated samples of M31 GCs\label{mc-tab}}
\tablehead{\colhead{Sample}&\colhead{$m^0_t$}&\colhead{${\sigma}_t$}}
\startdata
MP inner &16.17 &0.92\\
MP outer &17.00 &1.12\\
MR inner &15.91 &1.33\\
MR outer &16.35 &0.93\\
MP       &16.80 &1.13\\
MR       &16.28 &1.12\\
inner    &16.18 &1.08\\
outer    &16.74 &1.07\\
\enddata
\tablecomments{Uncertainties in the parameter values from the 200 individual trials are
comparable to those given in Table~\ref{all-tab}. Statistical uncertainties in
the averages are much smaller, typically 0.05~mag.}
\end{deluxetable}

\begin{deluxetable}{lrrrr}
\tablewidth{0pt}
\tablenum{6}
\tablecolumns{5}
\tablecaption{GCLF differences for full and de-contaminated samples of M31 GCs\label{diff-tab}}
\tablehead{\colhead{}&\multicolumn{2}{c}{Full samples}&\multicolumn{2}{c}{Decontaminated}\\
\colhead{Samples}&\colhead{$\Delta m^0_t$}&\colhead{$\Delta {\sigma}_t$}&\colhead{$\Delta m^0_t$}&\colhead{$\Delta {\sigma}_t$}}
\startdata
outer$-$inner      & 0.43& $-0.07$& 0.56&$ -0.01$\\
MP$-$MR            & 0.39& $-0.02$& 0.52&$  0.01$\\
MP outer$-$MP inner& 0.70& $ 0.08$& 0.83&$  0.20$\\
MR outer$-$MR inner& 0.31& $-0.58$& 0.44&$ -0.40$\\
MP outer$-$MR outer& 0.56& $ 0.15$& 0.65&$  0.19$\\
MP inner$-$MR inner& 0.17& $-0.51$& 0.26&$ -0.41$\\ 
\enddata
\end{deluxetable}

\clearpage


\end{document}